\documentclass[runningheads]{llncs}
\usepackage{times}
\usepackage[margin=1in]{geometry}
\usepackage{graphicx}
\usepackage{amsmath,amssymb}
\usepackage{float}
\usepackage{booktabs}
\usepackage{caption}
\usepackage{subcaption}
\usepackage{amsfonts}
\usepackage{url}
\usepackage{xcolor}
\usepackage{listings}
\usepackage{minted}
\usepackage{pifont}
\usepackage{booktabs}
\usepackage{tabularx}
\usepackage{array}
\usepackage{graphicx}
\usepackage{tikz}

\usepackage{xurl}
\usepackage{hyperref}

\usepackage{orcidlink}
\hypersetup{hidelinks}

\usetikzlibrary{positioning,arrows.meta,fit}

\begin{document}
%

\title{\textbf{RWA-PoB: A Credential-Based Proof-of-Backing Framework for Tokenized U.S. Treasury Products}}

\author{
Rischan Mafrur\inst{1,2}\orcidID{0000-0003-4424-3736}
\thanks{Corresponding author.}
\and
Gun Gun Febrianza\inst{3}
\and
Sean Foley\inst{4}
}
\authorrunning{R. Mafrur et al.}
%
\institute{
School of Business and Economics,
Universiti Brunei Darussalam, Brunei Darussalam\\
\email{rischan.mafrur@ubd.edu.bn}
\and
School of Computer, Data and Mathematical Sciences,
Western Sydney University, Australia
\and
Adaptive Institute, Indonesia\\
\email{gungunfebrianza@adaptive-institute.id}
\and
Department of Applied Finance,
Macquarie Business School,
Macquarie University, Australia\\
\email{sean.foley@mq.edu.au}
}

\maketitle              
\begin{abstract}
Proof of reserves (PoR) can improve transparency for tokenized assets, but
aggregate reserve coverage does not establish whether off-chain assets are
legally eligible, unencumbered, consistently valued, or sufficiently liquid for
redemptions. We propose RWA-PoB, a credential-based proof-of-backing framework
for tokenized U.S. Treasury products. Five authorised institutional roles
approve a canonical EIP-712 snapshot containing reserve, liability, liquidity,
and policy information. The framework evaluates backing adequacy through the
Backing Coverage Ratio (BCR) and short-term redemption capacity through the
Redemption Liquidity Coverage (RLC). The Solidity prototype couples the policy controller to an ERC-20 token. Successful issuance atomically increases token supply and recorded liabilities
by the corresponding USD-denominated liability. A redemption request burns tokens while reclassifying the corresponding obligation as pending. The obligation is reduced only after an
authorised settlement-role account confirms payment. We evaluate the framework
against a simplified aggregate PoR baseline using USDY-calibrated liabilities
and deterministic synthetic reserve scenarios. Both approaches permit issuance
in the valid state, but RWA-PoB rejects an encumbered-assets state with a BCR of
96.3408\%, below the experimental 105\% threshold. Under liquidity stress, it
classifies the proposed redemption as queued because post-request RLC falls to
39.9999\%. RWA-PoB authenticates the attribution and integrity of institutional claims, it
does not independently prove the existence, ownership, or condition of
off-chain assets. The prototype, test suite, datasets,
and replication scripts are available at
\url{https://github.com/rischanlab/PoB}.

\end{abstract}
\section{Introduction}
\label{sec:introduction}

Tokenization enables financial instruments and their associated ownership records to be represented and transferred through distributed ledgers. This architecture can improve automation and transparency in financial operations. It can also support programmable transactions, more efficient collateral use, and wider access to financial data. These benefits, however, are not automatic. Legal finality and asset custody continue to depend on institutions outside the blockchain. The same applies to valuation, redemption funding, and cash settlement.

Recent initiatives illustrate this relationship between on-chain systems and conventional financial infrastructure. BlackRock's BUIDL is a tokenized investment fund issued on public blockchain infrastructure. Swift, UBS Asset Management, and Chainlink have demonstrated the coordination of subscriptions and redemptions for tokenized funds through existing payment systems. DTCC's Smart NAV pilot has also examined the delivery of structured net asset value data to distributed-ledger applications \cite{securitize2024buidl,swift2024tokenizedfunds,dtcc2024smartnav}. Together, these initiatives demonstrate growing institutional adoption. They also show that tokenized products remain dependent on off-chain financial, operational, and legal processes.

The crypto-native concept of \emph{proof of reserves} (PoR) is commonly used to provide evidence that an issuer or custodian holds assets corresponding to its reported obligations. Depending on its design, a PoR system may disclose wallet balances, aggregate reserve values, custodian records, or independently verified attestations. This evidence can improve transparency, but an aggregate comparison between assets and liabilities does not by itself establish solvency or liquidity. It also does not establish whether token holders have enforceable claims over the reported assets.

Vidal-Tomás distinguishes among proof of assets, proof of liabilities, proof of reserves, and proof of solvency. He argues that reserve composition and omitted obligations can materially affect the interpretation of an apparently sufficient reserve balance \cite{vidaltomas2025}. This problem is broader for real-world assets (RWAs). The existence of an off-chain asset does not necessarily establish that it is legally owned by the appropriate entity or available to token holders. The asset may also be encumbered, unsettled, incorrectly valued, or unavailable within the promised redemption period.

These limitations are particularly important for tokenized securities. As emphasised by the U.S. Securities and Exchange Commission, placing a security on a blockchain does not change the legal nature of the instrument. The same substantive legal requirements continue to apply to its on-chain representation \cite{peirce2025tokenization}. Consequently, a technically valid token and an accurately reported on-chain supply do not independently establish the rights of token holders over the corresponding off-chain assets. A credible backing framework must therefore connect the on-chain state to evidence about custody and legal ownership. It must also account for segregation, encumbrance, valuation, settlement finality, and outstanding liabilities.

Existing systems address several parts of this problem. Chainlink Proof of Reserve can deliver off-chain or cross-chain reserve data to smart contracts. It can also be integrated with minting controls and automated circuit breakers \cite{chainlink2025por}. Treasury-backed products such as Ondo USDY publish daily third-party reserve attestations. They also report underlying asset values, collateralisation levels, and portfolio maturity information \cite{ondo2026usdy}.

The remaining problem is therefore not simply a lack of reserve disclosure or automated mint controls. Publicly documented systems do not generally define a common evidence model designed specifically for tokenized Treasury products. In particular, they do not consistently bind claims about custody, legal eligibility, encumbrance, valuation, settlement, liabilities, and redemption liquidity to the same reporting state. Without a common time reference and policy version, individually valid claims may describe different reserve conditions. This makes it difficult for a smart contract to determine whether the combined evidence supports a particular minting or redemption decision.

To address this problem, we propose \emph{RWA-PoB}, a credential-based proof-of-backing framework for tokenized U.S. Treasury products. RWA-PoB is not intended to provide a trustless cryptographic proof of real-world truth. 
RWA-PoB instead provides an attributable and machine-verifiable method for evaluating institutional claims under explicit trust assumptions. These claims are issued by identified custodians, administrators, valuation agents, legal or trustee roles, and independent verifiers. Once accepted, the claims are translated into on-chain controls for minting and redemption.

The contributions of this paper are as follows:

\begin{itemize}

\item We define a role-separated evidence model and a canonical reserve
snapshot for tokenized U.S. Treasury products. The model assigns institutional
responsibilities for custody, valuation, legal eligibility, liabilities, and
settlement, while binding the reported aggregates and reserve commitment to a
common epoch, valuation time, and policy version.

\item We distinguish backing adequacy from redemption liquidity through two
policy metrics. The \emph{Backing Coverage Ratio} (BCR) compares
risk-adjusted eligible reserves with total economic liabilities, whereas the
\emph{Redemption Liquidity Coverage} (RLC) evaluates whether liquid assets are
sufficient for redemption obligations under the applicable liquidity horizon.

\item We implement an on-chain policy controller that verifies five required
institutional signers, EIP-712 signatures, epoch progression, freshness,
expiry, policy consistency, liability continuity, and replay protection. The
accepted snapshot determines whether additional issuance satisfies the BCR and
RLC thresholds and whether a proposed redemption receives an instant or queued
disposition.

\item We integrate the policy controller with a non-upgradeable ERC-20 token so
that supply and liability changes occur atomically. Successful issuance
increases token supply and records the corresponding USD-denominated
liability. A redemption request burns tokens while reclassifying the
corresponding liability as a pending-redemption obligation, which is reduced
only after settlement is confirmed by an authorised role. The reported
prototype evaluates the special case $P_t^{\mathrm{red}}=1$.

\item We evaluate the prototype against a simplified aggregate PoR baseline
using USDY-calibrated liabilities and deterministic synthetic reserve
scenarios. The evaluation covers valid, encumbered-assets, and
liquidity-stress states, together with controller tests, adversarial
controller-token integration tests, a seeded accounting-invariant state
machine, and local EVM gas measurements.

\end{itemize}

RWA-PoB is not intended to replace regulated custodians, administrators, auditors, or legal agreements. Its purpose is to make their relevant claims attributable, consistently structured, time-bounded, and directly usable by smart-contract policy. The framework therefore enables an on-chain system to evaluate more than the reported existence of reserve assets. It can also assess whether those assets are eligible, sufficiently liquid, and adequate to support the complete set of token-related obligations under the stated assumptions.

\section{Literature Review}
\label{sec:literature-review}

\paragraph{Proof of Reserves, Liabilities, and Solvency.}
Research on reserve assurance initially developed in the context of
cryptocurrency exchanges and custodial platforms. Early cryptographic
approaches, such as Provisions, showed that an exchange could provide
privacy-preserving evidence that its controlled reserve assets exceeded the
liabilities included in a committed customer-balance set
\cite{dagher2015provisions}. Such mechanisms improve verifiability without
disclosing individual balances, while related accounting research examines
blockchain's potential to support more timely and verifiable assurance
processes \cite{dai2017blockchain}. However, cryptographic and ledger-based verification alone cannot guarantee the completeness of reported liabilities, the absence of competing claims on reserve assets, or the availability of sufficient liquidity to meet withdrawals when due.

Vidal-Tomás distinguishes among proof of assets, proof of liabilities, proof of
reserves, and proof of solvency \cite{vidaltomas2025}. The study shows that an
aggregate reserve snapshot should not be treated as a complete measure of
financial resilience. Its estimated reserve buffer of approximately 6--14\%
depends on the assets, institutions, and stress assumptions examined and should
not be applied directly to tokenized Treasury products. The broader finding is
nevertheless relevant: nominal reserve coverage may be misleading when asset
quality, market risk, omitted liabilities, and liquidation capacity are not
considered. Chainlink similarly notes that proof of assets and proof of
liabilities do not independently establish solvency, particularly when assets
may be pledged elsewhere or liabilities have different priorities
\cite{chainlink2023solvency}.

Lazirko \emph{et al.} address the separation between blockchain records and
conventional accounting information through their Double-Helix framework
\cite{lazirko2025doublehelix}. The framework distinguishes on-chain, off-chain,
and intersecting transactions and argues that reserve assurance must reconcile
both information environments. This is important because off-chain liabilities,
expenses, investments, and transfers may affect an entity's financial position
without appearing directly on the blockchain. However, Double-Helix primary focus is the financial position of cryptocurrency exchanges and custodial platforms not the backing of a specific tokenized security. It therefore does not define
how product-level claims concerning custody, legal eligibility, encumbrance,
valuation, liabilities, and settlement should be assigned to responsible
institutions. 

RWA-PoB addresses this related but distinct problem for tokenized Treasury
products. It binds role-specific institutional claims to a common signed snapshot and uses the accepted state to support minting and redemption
decisions. In this respect, RWA-PoB extends integrated reserve assurance from
exchange-level accounting reconciliation to product-level backing and
liquidity controls.

\paragraph{RWA Tokenization and the On-Chain--Off-Chain State Problem.}
The limitations of aggregate reserve assurance become more significant when the underlying assets exist outside the blockchain. In RWA tokenization, the ledger records a digital representation whose legal and economic validity depends on external institutions and agreements. Luo \emph{et al.} examine RWA tokenization across legal, technical, and cryptoeconomic layers \cite{luo2026sok}. Their analysis highlights the need to maintain consistency between deterministic on-chain records and externally administered assets, rights, and obligations. Taxonomic research indicates that surveyed RWA implementations commonly use
hybrid architectures \cite{vella2026taxonomy}. Within these architectures,
smart contracts may govern token representation, transfers, compliance rules,
pricing, and redemption workflows, whereas legal ownership, custody, dispute
resolution, and reserve verification remain anchored in off-chain arrangements.
Related research on blockchain-based asset trading further shows that ledger
design can affect the structure and timing of settlement
\cite{chiu2019settlement}.

This separation creates related oracle and evidence problems. The oracle
literature identifies off-chain data interfaces as potential failure points
because they do not inherit the reliability properties of the underlying
blockchain \cite{lo2020oracles}. A smart contract can verify a signature, data
commitment, or Merkle proof, but cannot independently determine whether a
custodian holds a Treasury security, whether that position is encumbered,
whether a purchase has reached settlement finality, or whether the governing
agreement gives the issuer an enforceable interest in the asset. An oracle can
transmit these claims on-chain, but the reliability of the resulting policy
decision depends on their source, scope, timing, and governance. The design
problem therefore extends beyond transmitting an aggregate reserve value to
determining which institution is authorised to attest to each material fact
and how related claims are bound to a common reporting state.

\paragraph{Credential and Token Standards.}
Existing standards provide several components that can support an RWA assurance architecture, although they address different parts of the system. ERC-3643 provides identity-based permissioning and compliance controls for the issuance, holding, and transfer of regulated tokens \cite{erc3643standard}. ERC-7943 defines a common interface for administrative and compliance functions in tokenized real-world assets. These functions include transfer validation, freezing, and forced transfers \cite{erc7943}. Such standards support compliant token administration, but they do not define how reserve eligibility, backing adequacy, or redemption liquidity should be assessed.

The W3C Verifiable Credentials Data Model 2.0 defines a standard data model for
tamper-evident credentials and presentations \cite{w3c2025vc}. Information
systems research similarly characterises verifiable credentials as mechanisms
for machine-verifiable exchange among issuers, holders, and verifiers, subject
to interoperability and governance requirements
\cite{sedlmeir2021credentials}. In a proof-of-backing context, a credential
could represent a custodian-confirmed position, an administrator-calculated
liability, or a valuation-agent price. Operational use nevertheless requires
more than verifying a digital signature: the system must define authorised
roles, common snapshot identifiers, validity periods, revocation procedures,
and the responsibilities of each credential issuer.

Together, these standards provide useful identity, compliance, and evidence components. They do not, however, prescribe how institutional claims should be combined into a complete financial backing state.

\paragraph{Industry Context and Case-Study Calibration.}
Industry reserve-assurance systems demonstrate that reserve information can
support on-chain controls. Chainlink Proof of Reserve can obtain data from
off-chain custodians or other sources and publish the resulting values to smart
contracts \cite{chainlink2025por}. Its \emph{Secure Mint} pattern can prevent
additional issuance when reported collateral is insufficient relative to token
supply \cite{chainlink2025principles}.

Tokenized Treasury issuers also provide detailed reserve information. Ondo USDY,
for example, publishes daily third-party reserve attestations covering aggregate
asset values, USDY obligations, collateralisation, portfolio composition,
liquidation value, and maturity indicators \cite{ondo2026usdy}. Its reporting
methodology retains unfulfilled redemption requests within outstanding
obligations until the redemption process is completed. This paper uses USDY
only as the case-study product and as the calibration context for the controlled
evaluation. USDY is not used as evidence of a product-specific assurance gap.

Adjacent research on money market funds and fiat-reserve-backed tokens links
stability to reserve composition, redemption arrangements, and portfolio
liquidity \cite{schmidt2016runs,oefele2024stablecoins}. These findings motivate
treating backing adequacy and short-horizon redemption liquidity as separate
policy questions. The research gap addressed here is methodological. Although
the reviewed literature and standards address reserve disclosure, oracle
delivery, credential authentication, accounting reconciliation, and
regulated-token controls, they treat these as distinct concerns. Taken
together, they do not provide a unified specification that binds role-assigned
institutional evidence to a common reporting state, evaluates backing and
redemption liquidity separately, and couples the resulting decisions to atomic
token issuance and redemption-accounting transitions. RWA-PoB addresses this
integration problem.

\paragraph{Market Development and Regulatory Context.}

The growth of tokenized financial markets provides further motivation for
stronger assurance mechanisms. Industry forecasts identify public securities
and liquid collateral, including U.S. Treasury securities, as likely early
drivers of institutional tokenization. Citi's 2026 \emph{Tokenization 2030}
report estimates a base-case market of approximately \$5.5~trillion in
tokenized assets by 2030 \cite{citi2026tokenization}. This projection indicates
significant market interest, but it does not imply that tokenization
automatically creates liquidity \cite{mafrur2026tokenized}, legal certainty, or lower financial risk.
These outcomes remain dependent on the structure of the instrument, the quality
of the underlying assets, the reliability of intermediaries, and the
enforceability of investor rights.

In the United States, the regulatory treatment of a tokenized Treasury product
depends on the rights and obligations represented by the token. Depending on
its structure, the instrument may constitute a fund interest, a debt
instrument, a security entitlement, or another contractual claim. SEC
Commissioner Peirce has emphasised that blockchain technology does not alter
the legal nature of the underlying instrument
\cite{peirce2025tokenization}. Tokenized securities therefore remain subject
to applicable securities laws. An on-chain token record, reserve commitment,
or smart-contract control cannot replace valid issuance documents, lawful
distribution, compliant custody, and accurate disclosure \cite{yeung2019regulation}.

A similar principle applies in the European Union. Crypto-assets that qualify
as financial instruments are generally excluded from the Markets in
Crypto-Assets Regulation and remain subject to the applicable financial-services
framework, including MiFID~II \cite{eu2023mica}. Regulation (EU) 2022/858
establishes the Distributed Ledger Technology Pilot Regime. Under this regime,
authorised market infrastructures may experiment with the trading and
settlement of DLT-based financial instruments under specified conditions and
supervisory oversight \cite{eu2022dltpilot}. Tokenized Treasury products cannot
therefore be classified uniformly under MiCA solely because they are
tokenized. Their regulatory treatment depends on their legal rights,
obligations, and economic substance.

IOSCO similarly identifies implications for investor protection, market
integrity, custody, settlement, interoperability, governance, and operational
resilience \cite{iosco2025tokenization}. Its technology-neutral approach
confirms that the use of DLT does not remove the responsibilities normally
associated with issuing, holding, transferring, valuing, and settling financial
instruments. Although these regulatory materials do not prescribe a particular
proof-of-backing framework, they identify requirements that such a framework
should support. These include attributable responsibility, timely disclosure,
effective custody controls, operational resilience, and enforceable investor
rights.

\section{Comparison with Existing Systems}
\label{sec:comparison}

Table~\ref{tab:comparison} positions RWA-PoB relative to three distinct approaches: document-based reserve attestations, Chainlink Proof of Reserve infrastructure, and the publicly documented USDY reserve-assurance model. These categories should not be interpreted as directly interchangeable products. A conventional attestation is principally a reporting mechanism; Chainlink Proof of Reserve is an oracle and enforcement infrastructure that can be integrated into different token systems; USDY is an operating tokenized Treasury product with its own legal, custody, verification, and disclosure arrangements; and RWA-PoB is the evidence and policy framework proposed in this paper.

\paragraph{Document-Based Attestations.}
Periodic audit or attestation reports can provide independently reviewed evidence concerning assets and liabilities at a specified reporting time. Their scope varies considerably. Some reports provide only aggregate balances, whereas others include asset categories, maturity information, custody arrangements, or reconciliation procedures. Such reports improve transparency and accountability but are normally consumed by investors, administrators, and regulators as documents. They do not inherently provide a machine-readable state that a smart contract can verify before executing an issuance or redemption action.

\paragraph{Chainlink Proof of Reserve.}
Chainlink Proof of Reserve provides infrastructure for obtaining reserve information from off-chain or cross-chain sources and publishing it to smart contracts \cite{chainlink2025por}. Its \emph{Secure Mint} pattern allows an issuing contract to compare reported reserves with token supply and to block additional issuance when collateral is insufficient. Chainlink can therefore provide both the data-delivery layer and an automated circuit-breaker mechanism.

The assurance offered by a particular Chainlink deployment nevertheless depends on the data sources, reporting methodology, oracle configuration, update policy, and smart-contract integration selected by the implementing project. Chainlink infrastructure does not, by itself, prescribe which institutional role must attest to legal ownership, custody segregation, encumbrance, settlement finality, or redemption liquidity. RWA-PoB is therefore complementary to an oracle network: a Chainlink feed or workflow could transport an RWA-PoB snapshot, while the RWA-PoB specification would define the evidence semantics, mandatory roles, financial metrics, and policy conditions applied to that snapshot.

\paragraph{Current USDY Reserve Assurance.}
USDY represents a stronger industry baseline than a simple aggregate PoR model. Ondo publicly describes USDY as debt issued by a bankruptcy-remote entity and backed by a portfolio that includes short-duration U.S. Treasury-related assets, bank deposits, and other eligible holdings under its governing arrangements \cite{ondo2026usdy}. Its public transparency process includes daily third-party reserve attestations, more detailed periodic reporting, aggregate underlying asset values, USDY obligations, collateralization information, liquidation value, portfolio composition, and maturity indicators.

USDY also incorporates legal and institutional protections. Public documentation identifies a verification and collateral-agent function, asset-eligibility restrictions, and a security interest intended to protect USDY holders \cite{ondo2026usdy}. Moreover, USDY's published reporting methodology does not remove a redemption request from outstanding obligations until the request has been fulfilled. RWA-PoB therefore does not claim to originate the accounting principle that unsettled redemptions remain liabilities.

\paragraph{RWA-PoB.}
RWA-PoB does not replace an auditor, custodian, collateral agent, oracle network, or token issuer. It defines how claims originating from these parties are transformed into one machine-verifiable backing state. The framework requires material claims to be attributable to authorised roles, bound to a common epoch and valuation cut-off, and checked for completeness, validity, freshness, revocation, and replay before they can affect token operations.

RWA-PoB further separates two financial questions that an aggregate collateralization ratio may combine or leave implicit. The Backing Coverage Ratio (BCR) evaluates whether risk-adjusted eligible reserves cover total economic obligations, whereas the Redemption Liquidity Coverage (RLC) evaluates whether sufficient assets can be converted to settlement assets within a specified redemption horizon. The accepted snapshot is then used by a smart-contract policy controller to determine whether issuance is permitted and whether a redemption can be settled immediately, queued, or subjected to predefined contingency procedures.

Table~\ref{tab:comparison} provides a descriptive comparison of capabilities
identifiable in publicly available sources. The comparison
distinguishes between disclosure, data delivery, and policy enforcement.
Chainlink provides oracle and reserve-linked control infrastructure, whereas
USDY combines third-party verification, legal arrangements, and public reserve
reporting. RWA-PoB contributes a common, role-separated evidence model that
binds these inputs to one time-consistent snapshot and makes their acceptance
an explicit condition for on-chain actions.

\begin{table}[t]
\centering
\caption{Descriptive comparison of capabilities identified in publicly
available sources.}
\label{tab:comparison}
\scriptsize
\setlength{\tabcolsep}{2.5pt}
\renewcommand{\arraystretch}{1.08}
\begin{tabularx}{\textwidth}{
    >{\raggedright\arraybackslash}p{0.29\textwidth}
    >{\centering\arraybackslash}X
    >{\centering\arraybackslash}X
    >{\centering\arraybackslash}X
    >{\centering\arraybackslash}X
}
\toprule
\textbf{Capability}
&
\textbf{Static attestation}
&
\textbf{Chainlink PoR}
&
\textbf{USDY}
&
\textbf{RWA-PoB}
\\
\midrule

Independent reserve evidence
&
Yes
&
Source-dependent
&
Yes
&
Required
\\

On-chain reserve-data delivery
&
No
&
Yes
&
Partial
&
Yes
\\

Reserve-linked mint control
&
No
&
Yes
&
Not publicly shown
&
Yes
\\

Role-separated claims
&
Report-dependent
&
Source-dependent
&
Partial
&
Yes
\\

Common time-consistent snapshot
&
No
&
Implementation-dependent
&
Not publicly shown
&
Yes
\\

Legal, custody, and encumbrance checks
&
Report-dependent
&
Source-dependent
&
Partial
&
Explicit
\\

Pending-redemption treatment
&
Report-dependent
&
Application-dependent
&
Yes
&
Explicit
\\

Separate solvency and liquidity metrics
&
Report-dependent
&
Application-dependent
&
Partial
&
BCR and RLC
\\

Credential freshness and replay protection
&
No
&
Implementation-dependent
&
Not publicly shown
&
Yes
\\

Programmatic policy enforcement
&
No
&
Supported
&
Partial
&
Yes
\\

\bottomrule
\end{tabularx}

\vspace{2pt}
\begin{minipage}{\textwidth}
\footnotesize
\end{minipage}
\end{table}

The comparison shows that RWA-PoB does not replace existing attestations,
oracle infrastructure, or issuer controls. Its contribution is to specify how
role-specific claims concerning reserves, liabilities, legal eligibility,
valuation, and settlement are combined into one canonical state. The framework
then verifies the completeness, freshness, and provenance of that state before
applying the BCR and RLC policy conditions to minting and redemption decisions.

\section{System Scope and Trust Model}
\label{sec:system-model}

This section defines the scope and operational assumptions of RWA-PoB, followed by the institutional trust relationships and threats considered in its design.

\subsection{Scope and Assumptions}
\label{sec:scope}

We consider fungible tokens representing economic claims on portfolios of
short-duration U.S. Treasury securities, cash, and cash equivalents held
through a fund, trust, or bankruptcy-remote special-purpose entity. The
conceptual framework includes Treasury bills and notes with remaining
maturities of up to two years; long-duration bonds, equities, and other RWA
classes are outside scope. The reported prototype experiments use
deterministically generated positions with maturities between one and 365
days.

The general framework values token obligations using an authenticated
redemption price or net asset value. In this framework, token quantities are
converted into USD-denominated liabilities using the redemption price included
in the accepted canonical snapshot. The reported prototype evaluates the
special case $P_t^{\mathrm{red}}=1$, under which token quantities and
USD-denominated liabilities are numerically equal and no variable-price
conversion is performed.

Prototype issuance is permissioned and may be initiated only by an account
holding the designated issuer role. Redemption requests, however, may be
initiated by any token holder with a sufficient balance. The prototype does
not implement investor whitelisting, transfer restrictions, or other
regulated-token compliance controls.

The framework assumes that institutional participants can provide digitally
signed, time-bounded evidence. In the prototype, this evidence consists of
EIP-712 signatures from five required institutional roles over the same
canonical snapshot. The controller verifies signer authorisation, signature
validity, timestamps, epoch progression, policy consistency, liability
continuity, and replay protection. It also permits administrative revocation
of the currently accepted snapshot. The prototype does not implement the
complete W3C Verifiable Credentials model, a credential-status list, or an
automated oracle or relayer. Such infrastructure may deliver signed snapshots
on-chain but is not treated as an independent source of real-world truth.

\subsection{Institutional Trust and Threat Model}
\label{sec:threat-model}

The framework distinguishes five roles that approve the canonical reserve
snapshot: custodian, administrator, valuation agent, legal or trustee role,
and independent verifier. A separate settlement role confirms partial or
final payment against recorded redemption claims. Token issuance is initiated
through an issuer role assigned to the integrated ERC-20 contract.

At the framework level, each institutional role is responsible for specified
claims. For example, the custodian is responsible for holdings and custody
status, the administrator for liabilities, the valuation agent for prices and
haircuts, and the legal or trustee role for ownership and asset eligibility.
Each of the five required signers signs the complete canonical
snapshot. The Solidity contract does not independently restrict a signer to a
particular subset of snapshot fields.

A valid digital signature establishes the origin and integrity of a submitted
snapshot, but not the truth of its contents. RWA-PoB therefore remains
dependent on institutional honesty, contractual obligations, regulatory
oversight, and legal enforcement. A materially false signed statement provides
attributable evidence of what an authorised account approved, but does not by
itself establish fraud, legal liability, or the existence of the reported
assets.

The threat model includes unauthorised or excessive issuance, omitted
liabilities, asset double-pledging, stale or conflicting evidence, replayed
snapshots, inconsistent reporting times, compromised signing keys,
unauthorised settlement confirmation, and manipulation or interruption of
snapshot delivery. Because acceptance requires one valid signature from each
of the five mandatory roles over the same digest, an honest required signer
can prevent acceptance of a state that it identifies as invalid. This
requirement improves safety but may reduce availability when a required signer
is unavailable or refuses to sign.

For the integrated token, the prototype addresses ordinary supply-accounting
bypasses by requiring issuance and the corresponding liability increase to
occur in one EVM transaction. A failed controller check reverts both state
changes. Redemption similarly couples token burning to the reclassification
of the same amount as a pending liability. These controls do not cover
separately issued wrapped or bridged representations, alternative token
contracts, or a malicious governance decision to authorise another liability
adapter.

The prototype treats the account holding the default administrative role as
trusted governance. This account can change the policy configuration, grant
or revoke institutional roles, authorise token adapters, and revoke the
current snapshot. Compromise or misuse of this authority remains outside the
ordinary signer and token-path protections. Production deployment would
therefore require additional governance controls such as multisignature
approval, timelocks, restricted token registration, and monitoring of role
changes.

The contract verifies authentication, signer authorisation, timestamp and
epoch conditions, policy consistency, liability continuity, and replay
protection. It does not independently verify the existence, ownership,
completeness, valuation, or legal status of off-chain assets. Similarly, a
settlement-role call records an authorised confirmation of payment but does
not cryptographically prove bank finality or transfer settlement assets.
Collusion among the required signers, undisclosed liabilities or legal claims,
custody failures, malicious governance, and incorrect but consistently signed
information remain outside the cryptographic guarantees of the framework.

\section{RWA-PoB Architecture and Contribution}
\label{sec:architecture}

USDY provides a useful case study because its publicly documented model includes
daily third-party reserve attestations, a redemption-price oracle, and
subscription and redemption functions \cite{ondo2026usdy,ondo2026contracts}.
USDY provides the product context and liability-scale calibration for the
controlled evaluation. 

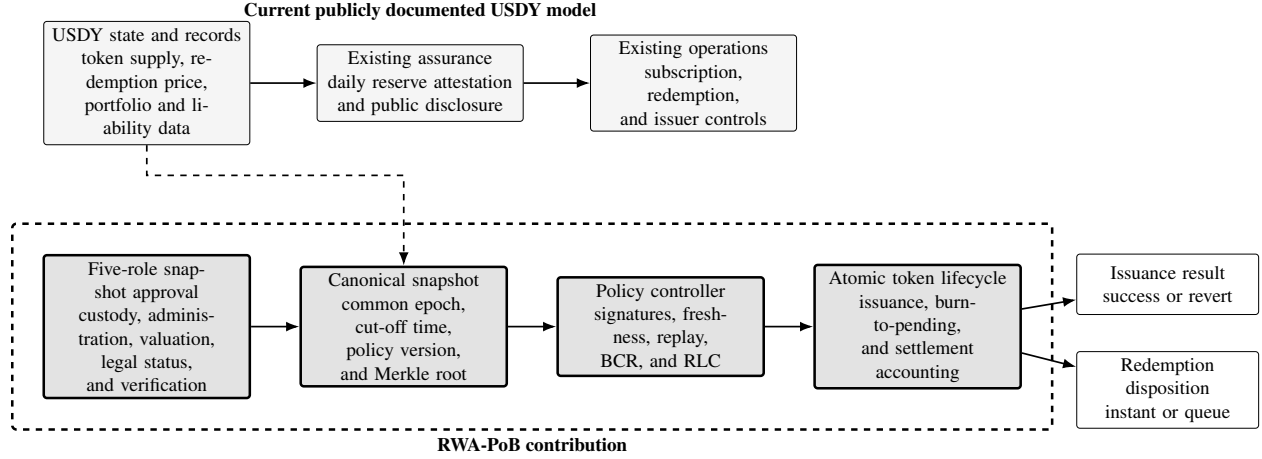
\begin{figure}[t]
\centering
\resizebox{\textwidth}{!}{%
\begin{tikzpicture}[
    font=\small,
    existing/.style={
        draw,
        rounded corners=1.5pt,
        fill=gray!8,
        align=center,
        minimum height=11mm,
        text width=3.2cm
    },
    proposed/.style={
        draw,
        very thick,
        rounded corners=1.5pt,
        fill=gray!22,
        align=center,
        minimum height=11mm,
        text width=3.2cm
    },
    output/.style={
        draw,
        rounded corners=1.5pt,
        align=center,
        minimum height=10mm,
        text width=2.8cm
    },
    flow/.style={
        -{Latex[length=2.2mm]},
        thick
    },
    note/.style={
        align=center,
        font=\footnotesize
    }
]

\node[existing] (usdydata)
{USDY state and records\\
token supply, redemption price,\\
portfolio and liability data};

\node[existing, right=11mm of usdydata] (attestation)
{Existing assurance\\
daily reserve attestation\\
and public disclosure};

\node[existing, right=11mm of attestation] (operations)
{Existing operations\\
subscription, redemption,\\
and issuer controls};

\draw[flow] (usdydata) -- (attestation);
\draw[flow] (attestation) -- (operations);

\node[note, above=3mm of attestation]
{\textbf{Current publicly documented USDY model}};

\node[proposed, below=18mm of usdydata] (evidence)
{Five-role snapshot approval\\
custody, administration, valuation,\\
legal status, and verification};

\node[proposed, right=8mm of evidence] (snapshot)
{Canonical snapshot\\
common epoch, cut-off time,\\
policy version, and Merkle root};

\node[proposed, right=8mm of snapshot] (controller)
{Policy controller\\
signatures, freshness, replay,\\
BCR, and RLC};

\node[proposed, right=8mm of controller] (token)
{Atomic token lifecycle\\
issuance, burn-to-pending,\\
and settlement accounting};

\node[output, right=9mm of token, yshift=7mm] (mint)
{Issuance result\\
success or revert};

\node[output, below=6mm of mint] (redeem)
{Redemption disposition\\
instant or queue};

\draw[flow] (evidence) -- (snapshot);
\draw[flow] (snapshot) -- (controller);
\draw[flow] (controller) -- (token);
\draw[flow] (token) -- (mint);
\draw[flow] (token) -- (redeem);

\draw[flow, dashed]
(usdydata.south) -- ++(0,-5mm) -| (snapshot.north);

\node[
    draw,
    very thick,
    dashed,
    rounded corners=2pt,
    fit=(evidence)(snapshot)(controller)(token),
    inner sep=5mm,
    label=below:{\textbf{RWA-PoB contribution}}
] {};

\end{tikzpicture}%
}
\caption{Existing USDY reserve assurance and the proposed RWA-PoB extension.
The contribution combines five-role snapshot approval, policy evaluation, and
atomic token-liability accounting; USDY provides the case-study environment.}
\label{fig:pob-architecture}
\end{figure}

RWA-PoB is proposed as an additional assurance layer.
Figure~\ref{fig:pob-architecture} distinguishes the existing assurance process
from the proposed contribution.

\paragraph{RWA-PoB contribution.}
The framework contributes four linked components:
\textbf{(i)} five-role snapshot approval;
\textbf{(ii)} a canonical backing snapshot;
\textbf{(iii)} a BCR- and RLC-based policy controller; and
\textbf{(iv)} atomic token-liability integration.
Together, these components convert a fresh, commonly approved backing state
into enforceable issuance and redemption-accounting transitions.

\subsection{Evidence and Institutional Roles}
\label{sec:evidence-model}

Table~\ref{tab:credential-roles} allocates institutional responsibilities
within the evidence model. Five roles approve the canonical snapshot, while a
separate operational role confirms subsequent redemption settlements.

\begin{table}[t]
\centering
\caption{Institutional roles in the RWA-PoB evidence model.}
\label{tab:credential-roles}
\footnotesize
\setlength{\tabcolsep}{4pt}
\renewcommand{\arraystretch}{1.10}
\begin{tabularx}{\textwidth}{
    >{\raggedright\arraybackslash}p{0.20\textwidth}
    >{\raggedright\arraybackslash}X
    >{\raggedright\arraybackslash}p{0.18\textwidth}
}
\toprule
\textbf{Role}
&
\textbf{Assigned responsibility}
&
\textbf{Function}
\\
\midrule

Custodian
&
Positions, quantities, custody, encumbrance, and settlement status.
&
Snapshot signer
\\

Administrator
&
Token liabilities, pending redemptions, fees, and other liabilities.
&
Snapshot signer
\\

Valuation agent
&
Asset values, valuation time, haircuts, and liquidity assumptions.
&
Snapshot signer
\\

Trustee or legal role
&
Legal ownership, asset eligibility, and applicable security interests.
&
Snapshot signer
\\

Independent verifier
&
Reconciliation of reserve, liability, and supporting records.
&
Snapshot signer
\\

Redemption agent
&
Confirmation of partial or final payment against a redemption claim.
&
Settlement role
\\

\bottomrule
\end{tabularx}
\end{table}

The custodian, administrator, valuation agent, legal or trustee role, and
independent verifier each sign the complete canonical snapshot. The prototype
therefore enforces role-separated approval, not field-level signatures. The
contract verifies exactly one authorised signer for each required role, while
the allocation of responsibility in Table~\ref{tab:credential-roles} remains
an off-chain institutional requirement.

The settlement role is not a required snapshot signer. It confirms partial or
final payment against an existing redemption identifier. This logical
separation must be maintained through role assignment because the contract
does not prevent governance from assigning settlement and snapshot roles to
the same account. The token's issuer role is also separate from the
administrator role and is used only to initiate atomic issuance.

\subsection{Reserve and Liability Model}
\label{sec:financial-model}

\subsubsection{Eligible Reserves}

The general framework represents each direct reserve position $j$ as
\begin{equation}
RSV_{j,t}
=
\left\langle
id_j,\,
q_{j,t},\,
p_{j,t},\,
\tau_j,\,
\kappa_{j,t},\,
\omega_{j,t},\,
\eta_{j,t},\,
\sigma_{j,t},\,
d_{j,t}
\right\rangle ,
\label{eq:reserve-state-vector}
\end{equation}
where $id_j$ is the unique identifier of reserve position $j$,
$q_{j,t}$ and $p_{j,t}$ denote its quantity and price,
$\tau_j$ is its maturity date, $\kappa_{j,t}$ identifies the custody account,
$\omega_{j,t}$ represents legal ownership, $\eta_{j,t}$ records encumbrance,
$\sigma_{j,t}$ records settlement status, and $d_{j,t}$ is the estimated
cash-conversion horizon.

A position is eligible only when the required legal and operational conditions
are satisfied:
\begin{equation}
\alpha_{j,t}
=
\mathbf{1}
\left\{
\begin{array}{l}
\text{custody confirmed, legal title valid,}\\
\text{no prohibited encumbrance, and settlement final}
\end{array}
\right\}.
\label{eq:eligibility-indicator}
\end{equation}

Its eligible value is
\begin{equation}
V_{j,t}^{\mathrm{eligible}}
=
\alpha_{j,t}
(1-h_{j,t})
p_{j,t}q_{j,t},
\label{eq:eligible-position-value}
\end{equation}
where $h_{j,t}$ is an asset- and horizon-specific haircut. The reported
experiment uses simplified synthetic records containing market, eligible, and
liquid values, maturity, haircut, encumbrance, legal eligibility, and
settlement-finality fields. Position eligibility and aggregation are computed
off-chain.

Total eligible reserves are
\begin{equation}
A_t^{\mathrm{eligible}}
=
A_t^{\mathrm{cash}}
+
\sum_j V_{j,t}^{\mathrm{eligible}}.
\label{eq:total-eligible-reserves}
\end{equation}

A Merkle root commits to the position records but does not prove their
completeness, truth, or consistency with the reported aggregates. The five
required signers approve the same root and aggregate values. On-chain
validation checks that the root is non-zero and that liquid assets do not
exceed eligible reserves, which do not exceed gross reserves.

\subsubsection{Economic Liabilities}

For a multi-network product, global economic supply is
\begin{equation}
S_t^{\mathrm{econ}}
=
\sum_{c\in\mathcal{C}} S_{c,t}
-
B_t^{\mathrm{locked}},
\label{eq:global-supply}
\end{equation}
where $B_t^{\mathrm{locked}}$ denotes tokens locked to support wrapped or
bridged representations included elsewhere in the summation. In-transit
amounts are treated under an explicit cross-chain policy. The prototype models
one token contract, for which $B_t^{\mathrm{locked}}=0$.

Token obligations are valued using the authenticated redemption price included
in the accepted canonical snapshot:
\begin{equation}
V_t^{\mathrm{tokens}}
=
S_t^{\mathrm{econ}}P_t^{\mathrm{red}}.
\label{eq:token-liability}
\end{equation}

This formulation captures the economic logic of an accumulating token such as
USDY, for which the redemption price may change over time. The reported
prototype and experiments evaluate the special case
$P_t^{\mathrm{red}}=1$. Consequently, token quantities and USD-denominated
liabilities are numerically equal in the reported implementation, although
they are distinct quantities in the general framework.

Under the prototype's burn-first redemption model, total liabilities are
\begin{equation}
L_t
=
V_t^{\mathrm{tokens}}
+
V_t^{\mathrm{pending}}
+
O_t,
\label{eq:total-liabilities}
\end{equation}
where $O_t$ includes fees and other liabilities, and
$V_t^{\mathrm{pending}}$ contains burned token claims that have not settled.
A lock-based implementation would need to avoid counting the same obligation
in both circulating supply and pending claims.

For a redemption request of $q$ token units, the corresponding liability is
$qP_t^{\mathrm{red}}$, using the authenticated redemption price when the
request is accepted. The request reduces $V_t^{\mathrm{tokens}}$ and increases
$V_t^{\mathrm{pending}}$ by the same USD-denominated amount, leaving total
liabilities unchanged.

When an authorised settlement-role account confirms payment of an amount $s$,
the prototype assumes that settlement is made from eligible cash. The
corresponding transition is
\begin{equation}
\begin{aligned}
V_{t+1}^{\mathrm{pending}}
&=
V_t^{\mathrm{pending}}-s,\\
A_{t+1}^{\mathrm{cash}}
&=
A_t^{\mathrm{cash}}-s,\\
A_{t+1}^{\mathrm{liquid}}
&=
A_t^{\mathrm{liquid}}-s,\\
A_{t+1}^{\mathrm{eligible}}
&=
A_t^{\mathrm{eligible}}-s,\\
A_{t+1}^{\mathrm{gross}}
&=
A_t^{\mathrm{gross}}-s,\\
L_{t+1}
&=
L_t-s.
\end{aligned}
\label{eq:settlement-accounting}
\end{equation}
These reductions are recorded as post-snapshot adjustments so that the
original signed snapshot remains unchanged. The settlement-role call records
an institutional confirmation of payment; it does not independently prove
bank finality.

Equation~\ref{eq:settlement-accounting} describes the complete framework-level accounting transition.
The reported prototype implements only the reductions in pending liabilities
and tracked liquid assets. It does not propagate settlement into cash,
eligible reserves, or gross reserves between snapshots.

\subsection{Backing Metrics and Policy Logic}
\label{sec:policy-logic}

RWA-PoB separates backing adequacy from short-horizon liquidity:
\begin{equation}
BCR_t
=
\frac{A_t^{\mathrm{eligible}}}{L_t},
\qquad
RLC_{t,h}
=
\frac{A_{t,h}^{\mathrm{liquid}}}
{R_{t,h}^{\mathrm{due}}},
\label{eq:bcr-rlc}
\end{equation}
where $A_{t,h}^{\mathrm{liquid}}$ is the eligible value convertible to the
settlement asset within horizon $h$, and $R_{t,h}^{\mathrm{due}}$ is the
redemption value due within that horizon. The metrics use the effective values
after post-snapshot issuance, redemption, and settlement adjustments. The
prototype uses one implicit horizon: current RLC divides liquid assets by
pending redemptions, while post-request RLC includes the proposed redemption
in the denominator. When no redemption value is due, the RLC condition is
satisfied by definition.

Snapshot acceptance requires
\begin{equation}
\operatorname{Accept}(\Sigma_t)
=
\operatorname{Roles}
\land
\operatorname{Signatures}
\land
\operatorname{Freshness}
\land
\operatorname{Epoch}
\land
\operatorname{Policy}
\land
\operatorname{Values}
\land
\operatorname{Continuity}
\land
\operatorname{UnusedDigest}.
\label{eq:snapshot-validity}
\end{equation}
Governance may subsequently revoke the accepted snapshot. A revoked, expired,
or policy-inconsistent snapshot is not fresh and cannot support issuance.

The policy controller applies the following rules:

\begin{enumerate}

\item \textbf{Snapshot update.}
A snapshot is accepted only when one authorised signer from each mandatory
role signs the same EIP-712 digest, its epoch increases, its timestamps and
values are valid, its token liability reconciles with the economic supply and
authenticated redemption price, its pending liability reconciles with unpaid
redemption claims, and its digest has not been used.

\item \textbf{Minting.}
The prototype assumes that subscription funds have settled and are already
included in eligible reserves. For a positive issuance of $q$ token units, let
$\Delta V_t(q)=qP_t^{\mathrm{red}}$ denote the corresponding USD-denominated
liability using the authenticated redemption price. Minting is permitted only
when the snapshot is fresh and
\begin{equation}
BCR_t^{\mathrm{post}}(q)
=
\frac{A_t^{\mathrm{eligible}}}
{L_t+\Delta V_t(q)}
\geq\theta_S
\quad\text{and}\quad
RLC_{t,h}\geq\theta_L.
\label{eq:mint-policy}
\end{equation}
Approval, the increase in token supply by $q$, and the increase in recorded
liabilities by $\Delta V_t(q)$ occur atomically.

\item \textbf{Redemption initiation.}
For a request of $q$ token units, the integrated token burns $q$ tokens and
atomically moves $\Delta V_t(q)$ from token liabilities to
pending-redemption liabilities. The applicable redemption price is fixed when
the request is accepted. If the controller rejects the request, the burn is
reverted.

\item \textbf{Redemption disposition.}
A request receives an instant disposition only when the snapshot is fresh and
post-request RLC, including $\Delta V_t(q)$ in the denominator, meets the
threshold. Otherwise, it receives a queue disposition. These values classify
the request but do not transfer assets or implement queue ordering.

\item \textbf{Settlement.}
An authorised settlement-role account confirms a positive partial or final
amount $s$ that does not exceed the unpaid claim or available liquid assets.
The reported prototype reduces pending liabilities and tracked liquid assets
by $s$ and closes the claim when fully settled. A complete implementation
should also update the cash, eligible-reserve, and gross-reserve aggregates
from which payment is made.

\end{enumerate}

A low BCR blocks further issuance, whereas a low RLC changes the redemption
disposition without extinguishing the investor's claim. The implemented flows
are shown in Appendix~\ref{app:policy-flows}.

\section{Prototype and Evaluation}
\label{sec:prototype-evaluation}

\subsection{Prototype Implementation}
\label{sec:prototype-implementation}

The prototype was implemented in Solidity~0.8.28 and evaluated using
Hardhat~3.11.1 on a local Ethereum-compatible network. It comprises three
deployed contracts and a narrow interface. The \texttt{PoBPolicyController} validates snapshots, calculates BCR and RLC,
authorises liability changes, records redemption dispositions, and accounts
for settlement. The reported implementation evaluates the special case
$P_t^{\mathrm{red}}=1$, so token quantities and USD-denominated liabilities
are numerically equal. Variable-price conversion and conversion rounding are
not implemented. The \texttt{RWABackedToken} atomically couples ERC-20
issuance and burning to the corresponding liability changes in the controller.
The \texttt{AggregatePoRBaseline} provides the gross-reserve comparison.

Each canonical snapshot is signed using EIP-712 by the custodian,
administrator, valuation agent, legal or trustee role, and independent
verifier. The controller checks signer roles and signatures, epoch and time
conditions, policy consistency, value bounds, liability reconciliation, and
replay protection. Under $P_t^{\mathrm{red}}=1$, token liabilities reconcile
directly with the registered token supply, while pending liabilities reconcile
with unpaid redemption claims. An accepted snapshot may subsequently be
revoked. Following confirmed settlement, the controller reduces effective pending
liabilities and liquid assets by the confirmed USD-denominated amount.
Because total liabilities are computed from token liabilities, pending
liabilities, and other liabilities, this reduction also decreases total
liabilities. The prototype does not update cash, eligible reserves, or gross
reserves following settlement.

The implementation, tests, scenarios, and reproduction scripts are available
at \url{https://github.com/rischanlab/PoB}.

\subsection{Experimental Design}
\label{sec:experimental-design}

The evaluation tests the policy layer and does not audit or reconstruct USDY's
reserve portfolio. Liability scale is calibrated using the RWA.xyz
\emph{Bridged Token Value (Dollar)} series for Ondo U.S. Dollar Yield~\cite{rwaxyz2026usdy}. The
processed dataset contains 1,043 daily observations from 18 September 2023 to
26 July 2026. The latest observation, approximately USD~2.162 billion, is used
in the experiments.

Synthetic position records contain market, eligible, and liquid values,
maturity, haircut, encumbrance, legal-eligibility, and settlement-finality
fields. Each scenario contains 100 positions generated with seed~42. Gross
reserves equal 115\% of observed token value, other liabilities equal 0.5\%,
and the proposed redemption equals 5\%. The experimental thresholds are
$\theta_S=105\%$ for BCR and $\theta_L=100\%$ for RLC.

Three scenarios are evaluated:

\begin{enumerate}

\item \textbf{Valid state:} eligible reserves exceed the BCR threshold and
liquid assets equal 10\% of observed token value.

\item \textbf{Encumbered-assets state:} gross reserves remain unchanged, but
encumbered positions are excluded from eligible reserves.

\item \textbf{Liquidity-stress state:} eligible reserves remain sufficient,
but liquid assets fall to 2\% of observed token value.

\end{enumerate}

The complete suite contains 31 deterministic tests: seven controller unit
tests, 21 atomic integration and adversarial tests, one seeded model-based test
of 128 state transitions, and two scenario and gas-generation tests. The suite
covers signature and snapshot validation, policy enforcement, role bypasses,
atomic rollback, redemption and settlement bounds, and accounting invariants.
All 31 tests passed in the post-merge continuous-integration run.

The USDY data are used to calibrate the dollar value of token liabilities.
The experiments do not reproduce USDY's historical token supply or changing
redemption price. Instead, they set $P_t^{\mathrm{red}}=1$ to isolate the
effects of reserve eligibility, backing coverage, redemption liquidity, and
liability reclassification.

\subsection{Experimental Results}
\label{sec:experimental-results}

Table~\ref{tab:scenario-results} reports the observed policy decisions. The
aggregate PoR baseline permitted minting in all three scenarios because gross
reserves exceeded token obligations. RWA-PoB agreed in the valid state but
rejected minting when encumbrance reduced eligible backing below the threshold.

\begin{table}[t]
\centering
\caption{Observed decisions in the USDY-calibrated scenarios.}
\label{tab:scenario-results}
\scriptsize
\setlength{\tabcolsep}{3.5pt}
\renewcommand{\arraystretch}{1.08}
\begin{tabularx}{\textwidth}{
    >{\raggedright\arraybackslash}X
    >{\centering\arraybackslash}p{0.16\textwidth}
    >{\centering\arraybackslash}p{0.16\textwidth}
    >{\centering\arraybackslash}p{0.14\textwidth}
    >{\centering\arraybackslash}p{0.10\textwidth}
    >{\centering\arraybackslash}p{0.12\textwidth}
}
\toprule
\textbf{Scenario}
& \textbf{Aggregate PoR mint}
& \textbf{RWA-PoB mint}
& \textbf{Disposition}
& \textbf{BCR}
& \textbf{Post-request RLC}
\\
\midrule

Valid state
& Accept
& Accept
& Instant
& 114.1209\%
& 199.9999\%
\\

Encumbered assets
& Accept
& Reject
& Instant
& 96.3408\%
& 199.9999\%
\\

Liquidity stress
& Accept
& Accept
& Queue
& 114.1209\%
& 39.9999\%
\\

\bottomrule
\end{tabularx}
\end{table}

In the encumbered-assets scenario, gross reserves were approximately
USD~2.486 billion against token obligations of approximately USD~2.162
billion. The aggregate baseline accepted minting, whereas RWA-PoB produced a
BCR of 96.3408\%, below the 105\% threshold, after excluding encumbered assets
and including other liabilities.

In the liquidity-stress scenario, BCR remained 114.1209\%, but liquid assets
covered approximately 40\% of the proposed redemption. The controller
therefore returned a queue disposition. This scenario test evaluates the
disposition without submitting a redemption; separate integration tests
confirm that a submitted request remains a pending liability until settlement.

\subsection{Execution Cost}
\label{sec:execution-cost}

Gas usage was obtained from local Hardhat transaction receipts.
Table~\ref{tab:gas-results} separates initial publication from subsequent
updates because the first transaction initializes empty storage.

\begin{table}[t]
\centering
\caption{Local EVM gas consumption for reserve-state publication.}
\label{tab:gas-results}
\small
\setlength{\tabcolsep}{5pt}
\renewcommand{\arraystretch}{1.08}
\begin{tabularx}{\textwidth}{
    >{\raggedright\arraybackslash}X
    >{\raggedleft\arraybackslash}p{0.22\textwidth}
    >{\raggedleft\arraybackslash}p{0.22\textwidth}
}
\toprule
\textbf{State publication}
& \textbf{RWA-PoB}
& \textbf{Aggregate PoR}
\\
\midrule

Initial valid state
& 362,241
& 94,689
\\

Encumbered-assets update
& 194,904
& 37,789
\\

Liquidity-stress update
& 197,716
& 37,789
\\

\bottomrule
\end{tabularx}
\end{table}

Initial RWA-PoB publication consumed approximately 3.83 times the baseline
cost, while subsequent updates consumed approximately 5.16--5.23 times the
baseline. The difference reflects five-signature verification and additional
state storage.

In ten controlled repetitions, atomic issuance used 132,573 gas. Separate
advisory authorisation and independent ERC-20 minting used 82,172 and 70,834
gas respectively, or 153,006 gas across two transactions. Atomic issuance was
therefore 20,433 gas, or 13.35\%, lower in this experiment while enforcing one
synchronised state transition.

Scenario inputs, Merkle roots, and policy decisions reproduce deterministically
under the recorded dependencies and seed. Gas values are run-specific because
timestamp-dependent EIP-712 signatures can change calldata composition. They
exclude deployment, off-chain processing, oracle fees, and network prices.

\section{Regulatory and Legal Considerations}
\label{sec:legal}

The legal treatment of a tokenized Treasury product depends on its structure
and the rights represented by the token. In the United States, it may
constitute a fund interest, debt security, security entitlement, or other
contractual claim. Distributed-ledger technology does not change the legal
character of an underlying security \cite{peirce2025tokenization}. Applicable
requirements may include registration or exemptions, disclosure, custody,
transfer restrictions, investor eligibility, and anti-money-laundering
controls.

In the European Union, MiCA does not apply to crypto-assets that qualify as
financial instruments, which remain subject to the applicable
securities-market framework. The DLT Pilot Regime establishes a supervised
framework for certain DLT-based market infrastructures
\cite{eu2023mica,eu2022dltpilot}. Product classification and jurisdiction must
therefore be determined before selecting legal, credential, and policy
requirements.

RWA-PoB does not determine or enforce legal compliance. The conceptual
framework could be combined with ERC-3643 or external identity and compliance
systems \cite{erc3643standard}. The implemented token, however, is a standard
ERC-20 with unrestricted transfers. It does not implement investor
whitelisting, identity verification, freezing, forced transfers,
jurisdictional restrictions, or anti-money-laundering controls.

A document hash or Merkle commitment establishes that presented data match a
commitment; it does not prove the data are true or create legal rights.
Enforceable rights must arise from applicable fund documents, security
agreements, custody arrangements, and law. In the framework, a position whose
ownership, custody, or enforceability cannot be established receives
$\alpha_{j,t}=0$ and is excluded from eligible reserves. The prototype does not
establish these legal facts independently; it relies on off-chain eligibility
classification and institutional approval of the signed snapshot.

\section{Limitations and Future Research}
\label{sec:limitations}

RWA-PoB authenticates the origin and integrity of signed institutional
statements, not their economic or legal truth. Signer collusion, omitted
liabilities, false custody or valuation records, and defective legal
arrangements may therefore remain undetected. The prototype enforces
role-authorised approval but does not independently establish organisational
independence among the signers or restrict each signer to specific snapshot
fields. A settlement-role call records an institutional confirmation of
payment but does not prove payment or bank finality. Similarly, a Merkle root
commits to position records but does not establish portfolio completeness or
consistency with the reported aggregates, and the controller neither verifies
position proofs nor recomputes reserve totals. Governance can change policies,
grant or revoke roles, authorise token contracts, and revoke the current
snapshot. Requiring all five roles to approve each snapshot improves safety
but creates an availability risk because one unavailable or withholding signer
can prevent an update, halt issuance, and cause redemptions to be queued.

The general framework supports a snapshot-authenticated redemption price,
whereas the reported prototype evaluates the special case
$P_t^{\mathrm{red}}=1$. The simplified PoR
baseline does not represent encumbrance or redemption liquidity, so the
comparison illustrates the effect of applying the additional eligibility and
liquidity conditions. The experiments use the latest observation from the
1,043-day liability series, synthetic positions generated with seed~42, one
deterministic state-machine sequence, and local EVM gas measurements. They do
not provide longitudinal simulation, real portfolio calibration, valuation or
availability sensitivity, scaled network-cost analysis, formal verification,
an independent security audit, or production validation.

Future work should formalise the adversary model and the safety and liveness
properties of the framework, including supply-liability reconciliation under
price updates, liability conservation during redemption initiation, and
reserve reduction following settlement. These properties should be evaluated
through fuzz-invariant testing, symbolic execution, and a
threat-to-outcome matrix. Longitudinal simulations should use the complete
liability series, stochastic redemption flows, and parameter sweeps to measure
threshold binding, queue duration, false pauses, and valuation sensitivity.
Further evaluation should use position-level regulatory or issuer data and
compare RWA-PoB with aggregate-reserve, eligible-reserve, and single-attestor
baselines. Future work should also quantify signer availability, examine
threshold-signature alternatives, calibrate policy parameters against
applicable requirements, and measure gas across signer counts, policy paths,
and deployment networks. Production studies should integrate live price and
credential-status infrastructure, verifiable settlement evidence, compliant
transfers, cross-chain supply reconciliation, and hardened governance.

\section{Conclusion}
\label{sec:conclusion}

Aggregate proof of reserves does not by itself establish that reported assets
are legally eligible, unencumbered, consistently valued, or available within a
required redemption horizon. This paper proposes RWA-PoB, a credential-based
assurance framework for tokenized U.S. Treasury products. It combines
five-role approval of a canonical snapshot with explicit backing, liquidity,
issuance, and redemption-accounting policies.

RWA-PoB makes four principal contributions. First, it verifies signer roles,
signatures, freshness, policy consistency, liability reconciliation, and replay
protection. Second, it separates backing adequacy from redemption liquidity
through BCR and RLC. Third, it defines the atomic coupling of ERC-20 issuance to the corresponding USD-denominated liability. The reported prototype evaluates this mechanism
under the special case $P_t^{\mathrm{red}}=1$. Redemption requests reclassify the same USD-denominated value as a pending obligation until settlement is confirmed. Fourth, it
provides reproducible scenario, invariant, adversarial, and gas evidence for
these controls.

In the controlled evaluation, RWA-PoB rejected an encumbered-assets state that
the aggregate PoR baseline accepted. Under liquidity stress, it classified the
proposed redemption as queued despite adequate overall backing. Integration
tests further showed that submitted redemptions remain pending liabilities
until settlement and that confirmed payments reduce both the outstanding liability and tracked liquid assets. These results demonstrate the
intended policy and accounting behaviour of the prototype but do not remove
institutional trust or establish production readiness. Broader empirical,
legal, operational, and security validation remains necessary.

\appendix

\setcounter{figure}{0}
\renewcommand{\thefigure}{\thesection.\arabic{figure}}

\section{RWA-PoB Policy-Flow Diagrams}
\label{app:policy-flows}

\begin{figure}[H]
    \centering
    \includegraphics[
        width=\textwidth,
        height=0.72\textheight,
        keepaspectratio
    ]{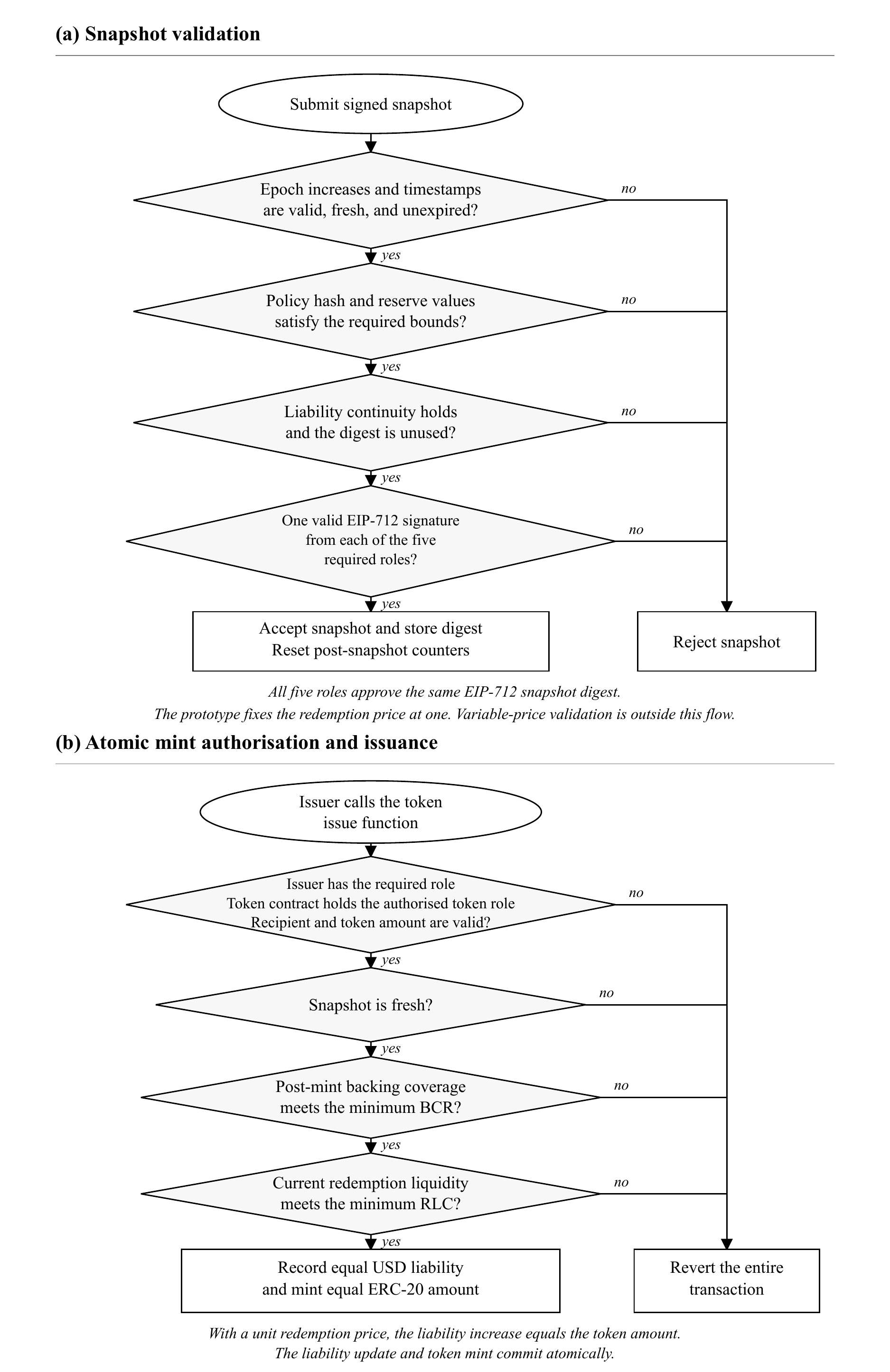}
    \caption{RWA-PoB snapshot-validation and mint-authorisation flows.}
    \label{fig:snapshot-mint-flow}
\end{figure}

\begin{figure}[H]
    \centering
    \includegraphics[
        width=\textwidth,
        height=0.82\textheight,
        keepaspectratio
    ]{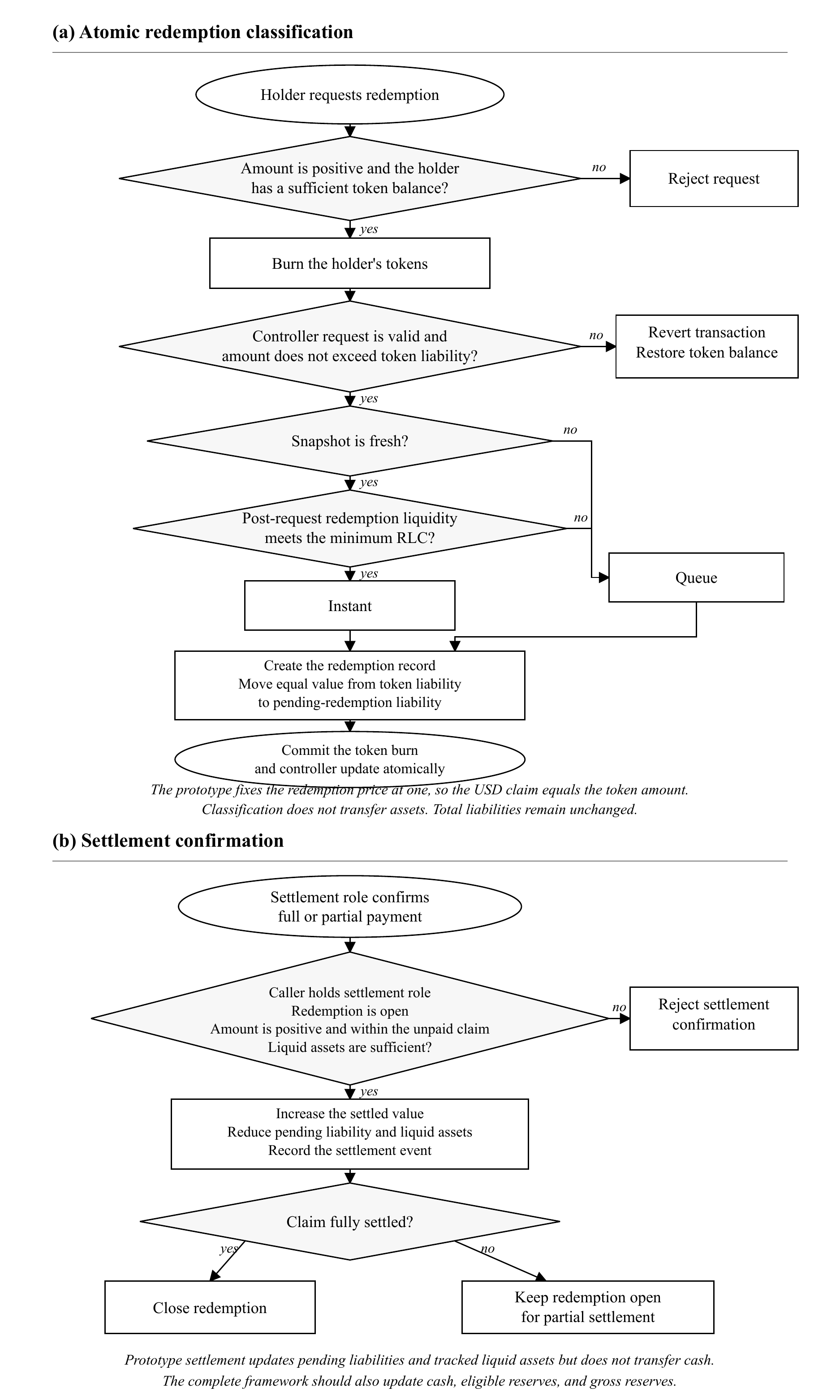}
 \caption{RWA-PoB redemption-classification and settlement-confirmation flows.
A redemption burns $q$ tokens while reclassifying the corresponding
request-time USD liability $\Delta V_t(q)$ as pending. In the reported
prototype, $P_t^{\mathrm{red}}=1$ and therefore $\Delta V_t(q)=q$.}
    \label{fig:redemption-settlement-flow}
\end{figure}

This appendix presents the operational decision flows implemented by the
RWA-PoB prototype. Figure~\ref{fig:snapshot-mint-flow} summarises snapshot
validation and mint authorisation. Figure~\ref{fig:redemption-settlement-flow}
shows redemption classification and subsequent settlement confirmation.

\clearpage

\end{document}